# Raman Scattering Investigation of Electron Phonon Coupling in Carbon substituted $MgB_2$

T.Sakuntala[#], A. Bharathi[*], S. K. Deb[#], N.Gayathri, C.S. Sundar and Y.Hariharan

Materials Science Division, Indira Gandhi Centre for Atomic Research, Kalpakkam - 603102, India
[#]Synchrotron Radiation Section, Bhabha Atomic Research Centre, Mumbai - 4000085, India

**Abstract**

Room temperature Raman scattering measurements have been carried out on well characterized samples of $MgB_{2-x}C_x$. The Raman line corresponding to the $E_{2g}$ phonon mode shows progressive hardening from 620 $cm^{-1}$ in pristine $MgB_2$ to 775 $cm^{-1}$ in the sample with carbon fraction x=0.2. The corresponding line width on the other hand, increases from a value of about 220 $cm^{-1}$ to 286 $cm^{-1}$ in samples with x = 0.1, beyond which it decreases to a value 167 $cm^{-1}$ for x=0.2. From the average mode frequency and the line width obtained from Raman measurements and taking the values of N(0) obtained from the calculated variation in σ hole density of states in $MgB_{2-x}C_x$, the electron phonon coupling strength to the $E_{2g}$ phonon, $\lambda_{2g}$, is evaluated using Allen's formula. This remains large for low C fraction, but shows rapid decrease for x > 0.10. Using this value of $\lambda_{2g}$ appropriately weighted, $T_C$ is obtained from McMillan's equation. These values are in good agreement with the experimentally measured $T_C$ variation in $MgB_{2-x}C_x$.

*corresponding author, bharathi@igcar.ernet.in
Dr. A. Bharathi,
Materials Science Division,
Indira Gandhi Centre for Atomic Research
Kalpakkam - 603102, India

# INTRODUCTION

The discovery of superconductivity at 39 K in $MgB_2$ [1] has attracted much attention not only because of the unusual two-gap nature of superconductivity in this system [2], but also because of the possibility of attaining high critical fields [3]. In this connection vast literature is now accruing on the increase in the critical fields consequent to C substitutions [4,5], with $H_{C2}$ as large as 32 T. The synthesis of C substituted $MgB_2$ was much debated. Consequent to several experimental investigations [6-8], there is now a consensus that the $T_C$ and the a-lattice parameter are well established functions of the carbon fraction. Further, the $T_C$ and a-lattice parameter show a definite correlation irrespective of the method of synthesis [9-11]. From these experimental results it is now clear that the $T_C$ of $MgB_2$ with ~10 at.% of the B substituted with carbon is 22 K. This decrease in $T_C$ to 22 K, is much smaller than that expected from rigid band behavior. Understanding the $T_C$ dependence and robustness of two gap superconductivity in carbon doped $MgB_2$ [12,13] has motivated several band structure calculations [14 -17]; in particular, the effect of electron doping on the σ band has been investigated in [16]. These calculations indicate that for x=0.2 in $MgB_{2-x}C_x$ (corresponding to 10 at.% C substitution), the σ band holes are still present and that for x=0.4 they completely vanish at $E_F$. It is also suggested that electron-phonon coupling strength mediated by the $E_{2g}$ bond-stretching mode should increase substantially [16] for small C substitutions.

This $E_{2g}$ mode is Raman active [18,19] and couples strongly with the σ band holes leading to a high $T_C$ of 39 K in $MgB_2$. Thus the role of Raman scattering in understanding superconductivity in $MgB_2$ is expected to be significant, and such studies have been extensive [20]. The $E_{2g}$ mode in $MgB_2$ appears at ~600 cm$^{-1}$, which is much lower than in the isostructural, non-superconducting $AlB_2$ sample wherein it appears at 980 cm$^{-1}$. This phenomenal softening of the $E_{2g}$ bond stretching mode and excessive broadening ( ~ 200 cm$^{-1}$) is understood as largely due to electron phonon coupling [21]. The $E_{2g}$ mode has also been investigated using inelastic x-ray scattering (IXS) [22], the results of which clearly indicate that the observed broadening 20-28 meV along Γ-A line, is primarily due to electron-phonon (e-p) coupling and not due to anharmonicity, since the magnitude of the latter is only ~1.2 meV at 300 K. Evolution of the Raman mode as a function of pressure up to 15 GPa in pure $MgB_2$ [23], indicated a large mode Gruneisen parameter. Raman scattering studies have also proven to be useful in studying the phonon behavior in substituted $MgB_2$. Systematic studies on $Mg_{1-x}Al_xB_2$ as a function of x revealed that the $E_{2g}$ phonon frequency softens from 980 $cm^{-1}$ in pure $AlB_2$ to 600 $cm^{-1}$ in $MgB_2$ and the corresponding line width increases from 40 $cm^{-1}$ to a value greater than 200 $cm^{-1}$ [24]. Such studies on C substituted $MgB_2$ are limited. Arvanitidis et al reported Raman studies in polycrystalline C substituted $MgB_2$ samples restricting to low concentration of x = 0.08 [25]. Lee et. al [26] reported Raman spectra for high carbon composition upto x = 0.25, in single crystalline form. In the present work we study the evolution of the $E_{2g}$ phonon with C substitution in $MgB_{2-x}C_x$ for a wide range of composition. Measurements have been carried out on well characterized samples [6,11]. From the observed frequency and the line width variations of the Raman shift, and the calculated σ hole density of states, the e-p coupling strength is evaluated as a function of carbon fraction. This variation of e-p coupling is suitably incorporated in the McMillan's equation for calculating $T_C$. The agreement between the measured $T_C$'s and those calculated as above is found to be satisfactory.

**EXPERIMENTAL**

The samples of $MgB_{2-x}C_x$ for x=0.0 to 0.06, used in the present investigation have been synthesized at 900 $^0C$, under 50 bar Ar pressure [6], and those with x > 0.06 were prepared by reaction of elemental constituents at 1250 $^0C$ in sealed Ta tubes [11] and were characterized by measuring the superconducting transition temperatures $T_C$ and lattice parameters [6]. The samples were phase pure and had sharp superconducting transitions. The a-lattice parameters were used to determine the exact C stoichiometry, using the procedure outlined in Ref [4]. $T_C$ was obtained from the temperature of the onset of diamagnetism. $T_C$ versus a-lattice parameter showed a correlation similar to that in single crystals [11]. Raman measurements were carried out on the as-prepared chunks. A 532 nm laser source of power 15 mW was used to excite the Raman line. Scattered light was detected using a CCD based (ANDOR technology) home built [27] Raman spectrometer with 600 lines/mm grating together with a super- notch filter covering the range of 200-1750 cm$^{-1}$. On each sample several measurements were carried out at different spots. Raman spectra obtained from different regions of the sample were similar implying a good homogeneity of the sample. The Raman spectrum could be fitted to a Gaussian / Lorenztian line shape. It was noted that for substitutions up to x=0.04, the Lorentzian fits were better, whereas for higher concentrations the fits to the Gaussian line shape was better. In the present work, all the spectra are fitted using a single Gaussian profile and the parameters from the fit, are used for further analysis.

## RESULTS

$T_C$, obtained as the onset of diamagnetic signal, for various carbon fractions x is shown in Fig.1. The observed $T_C$ dependence on x is in excellent agreement with those reported in literature [11]. The Raman spectra of some representative samples, after subtracting a linear background, are shown in Fig.2. The mode frequency and the linewidth in pristine $MgB_2$ in the present work are close to those reported by Renker et. al. [21] in polycrystalline samples. The line shape in the pristine sample, however is nearly symmetric. For higher carbon fraction, the average phonon frequency is obtained by fitting the profile to a single Gaussian, which is also shown as solid line in Fig. 2. It is clear from the figure that the peak position shifts to higher wave number with increasing C content and the width increases. The variations of the frequency and linewidth as a function of carbon fraction are shown in Figs.3a & 3b respectively. With increasing carbon fraction, the mode frequency shows an increase for concentrations up to x=0.05, and remains nearly constant over the range x = 0.06 - 0.10. Beyond x=0.1 the frequency increases continuously and has a value of 775 cm$^{-1}$ for x=0.2. The line width gradually increases from 220 cm$^{-1}$ to a maximum value of 286 cm$^{-1}$ at carbon fraction of x ~ 0.1 (Fig. 3b) beyond which it shows a precipitous drop to 167 cm$^{-1}$. This behavior is qualitatively similar to the earlier single crystal work, wherein Lee et. al [24] observed a broad peak for x=0.1, with peak position ~800 cm$^{-1}$, which narrows considerably for x=0.25 in the non-superconducting sample. Arvanitidis et al [25] reported that the spectra could be fitted to two components, one corresponding to the E$_{2g}$ phonon and other due to the peak in the phonon density of states. While the former shows broadening followed by a narrowing with increasing C substitution, the peak due to phonon DOS shows a monotonic increase in width, with C substitution. Thus, considering the overall linewidth dependence on C doping, an initial broadening followed by narrowing is clearly noted in the present work as well as the earlier work [24,25]. This is qualitatively similar to the behavior in Mg$_{1-x}$Al$_x$B$_2$, wherein line broadening was noted for increasing Al content up to x=0.3 [21], beyond which there is considerable narrowing.

At higher carbon compositions, typically above x = 0.1, Raman spectrum is noted to have relatively sharp bands around 640, 696 and 740 cm$^{-1}$ riding on a broad profile. Position of these peaks are very close to those in the calculated phonon density of states reported by Osbourn et al [28]. Further, impurity phases like MgB$_2$C$_2$ lead to strong intensities in the range 1000-1300 cm$^{-1}$, absence of which suggests that these peaks are rather due to phonon density of states appearing in the Raman spectrum due to increasing disorder and not from impurity phases.

## DISCUSSION

Increase in the phonon frequency due to carbon doping has contributions from a decrease in the a-lattice parameter and changes from altered e-p coupling. Variation in mode frequency arising solely from change in a-lattice parameter (brought about by the application of pressure) is shown as solid line in Fig. 3a. This accounts for an increase in the mode frequency from 620 to 675 cm$^{-1}$ for x = 0.2. Corresponding increase in line width expected is about 10 cm$^{-1}$. These values are small compared to the changes observed due to carbon substitution, as shown in Figs.3a & b, implying that contribution to the changes in γ and ω arise largely due to changes in the electronic structure, viz., due to electron doping rather than due to lattice contraction.

As mentioned in the introduction, it is now widely accepted that the significant softening of the in-plane bond stretching frequency in MgB$_2$ is attributed to the strong electron - phonon coupling and consequently, to a high value of the transition temperature T$_C$. Conversely, an increase in E$_{2g}$ phonon frequency is believed to be correlated to a reduced T$_C$ [25,26]. A small, but definite increase in T$_C$ compared to the bulk in strained MgB$_2$ films grown on SiC substrate [29] is understood along similar lines. Increase in the phonon frequency in carbon-substituted samples in the earlier work [25,26], were also attributed to a reduced electron-phonon coupling and in-turn to a decreased T$_C$. However, one has to consider the dependence of the e-p coupling strength λ on x in order to understand the T$_C$ dependence on x. Taking cue from IXS analysis, Allen's formula [30] is used to relate the line width, γ to the Raman shift, ω, viz.,

$$\gamma = 2\pi\lambda_{2g} N(0)\omega^2 \qquad (1)$$

where N(0) is the total density of states at E$_F$ and λ$_{2g}$ is the electron-E$_{2g}$ phonon coupling strength. The quantity γ/ω$^2$, obtained from the present work, is plotted in Fig.4, as a function of the carbon fraction x. It can be seen from the figure that γ/ω$^2$, does not show any significant change over the range x = 0.0 to 0.10, remaining at ~ 4.5 eV$^{-1}$; but beyond x=0.1, it drops to ~2.0 eV$^{-1}$. The calculated variation of the number of holes in the sigma band [16] as a function of carbon content, assuming a linear interpolation to be valid for intermediate compositions, is also shown in Fig.4 as solid line.

For pristine MgB$_2$ the total DOS at E$_F$ is taken as N(0)=0.354 states/ eV-fu-spin, with contribution from σ band being 0.15states/eV-fu-spin and that from the π band being 0.204states/eV-fu-spin[17,32]. Upon C substitution it is assumed that the π band contribution to the hole DOS N(0) remains constant and that only the σ band contribution to N(0) varies. Taking the cue from the calculation [16] it is argued that the σ hole DOS is proportional to the number of holes in the σ band. The latter has been calculated for x=0 and x=0.167 using a supercell method [16]. Using a linear interpolation to obtain the σ band contribution to N(0) as a function of C content, the solid line shown in Fig.4 is obtained. Adding to this a constant term due to the π band (0.204 states/eV-fu-spin), we obtain the dependence of N(0) on x.

Knowing γ/ω$^2$ and N(0) as a function of x, λ$_{2g}$ is evaluated using Eqn. 1. λ$_{2g}$ thus obtained is shown in Fig.5, from which it is clear the electron-phonon coupling strength remains high for small C substitutions up to x=0.1 beyond which it shows a precipitous fall. It is noteworthy that the value of λ$_{2g}$ obtained for pristine MgB$_2$ viz., 2.04±0.4, is very close to 2.5±1.1 obtained for the q=0.2Γ-A, E$_{2g}$ phonon from inelastic x-ray scattering measurements [22]. It may be mentioned that the calculations indicated that λ$_{2g}$ would diverge for low carbon substitution, which is however not noted in Fig. 5. Further, the calculations [16] suggested that there would not be significant shift expected in the phonon frequency with C substitution. On the other hand, experimental results clearly indicate that there is indeed a significant shift in the peak position, which cannot be fully accounted for from lattice contraction alone, implying that the reduction in hole density does affect the peak-position in this system.

It is well known that apart from the electron coupling to the E$_{2g}$ phonon, determined from Allen's formula, the λ$_{tot}$ that enters McMillan Equation has contributions from other phonons as well. So, λ$_{tot}$ arises from E$_{2g}$ phonons and other phonons, viz., λ$_{tot}$ =λ$_{2g}$+λ$_{other}$. Further to obtain reasonable values of T$_C$, λ values from E$_{2g}$ and other phonons in pristine MgB$_2$ is restricted to be 0.4, in accordance with Pickett et al [31]. Scaling the $\lambda_{2g}$ for all C fraction results in the evaluation of $\lambda_{tot}$ as a function of C content. Using this value of $\lambda_{tot}$ and $\mu^*=0.1$ in the McMillan's equation, $T_C$ is evaluated for various x and is shown in Fig. 6. The experimentally measured $T_C$ is also shown in Fig. 6 for comparison. A large increase in the calculated $T_C$ for x=0.02 sample is a direct consequence of the rather large linewidth at this composition.

It is also possible to obtain $\lambda_{2g}$ from the well known Hopfield expression $\lambda_{2g}=N_\sigma(0)<I^2>/M\omega^2$, where $N_\sigma(0)$ is the σ-band hole DOS at $E_F$, $<I^2>$, the electron-phonon matrix element averaged over the Fermi surface, M the atomic mass and ω, the phonon frequency corresponding to the $E_{2g}$ mode. Taking into account the variation of the σ-band hole DOS at $E_F$ and the $E_{2g}$ phonon frequency variation as a function of carbon content x from Fig.3, it can be seen from calculations similar to that described earlier, that $T_C$ decreases from 39 K to 14 K for x = 0.2, a reduction much larger than that observed in experiments, as shown in Fig.6. On the other hand, $\lambda_{2g}$, obtained using Allen's formula which incorporates changes in both the line width and the frequency leads to a $T_C$ variation, closer to that observed in $MgB_{2-x}C_x$.

## SUMMARY

Raman scattering measurements were carried out at room temperature in $MgB_{2-x}C_x$ samples for x=0.0 to x=0.2. The average phonon frequency increases with C substitution, whereas the linewidth increases to a maximum for x=0.1 beyond which there is a decrease. The electron-$E_{2g}$ phonon interaction parameter $\lambda_{2g}$ is extracted using Allen's formula [29]. The value for $\lambda_{2g}$ thus obtained in pristine $MgB_2$ is in good agreement with that obtained from X-ray inelastic scattering [24]. By suitably incorporating the $\lambda_{2g}(x)$ variation in $\lambda_{tot}$ appearing in McMillan's equation, the variation of $T_C(x)$ is calculated, which compares well with the measured $T_C$ in $MgB_{2-x}C_x$.

**Figure Caption**

Fig. 1 The superconducting transition temperature $T_C$ as function of the actual carbon fraction, as obtained from $\Delta a$-x correlation of Ref [4].

Fig. 2 The Raman spectra in $MgB_{2-x}C_x$ for the various carbon fraction x. Solid lines are fits to Gaussians.

Fig.3 (a) Variation phonon frequency $\omega$, with C fraction x in $MgB_{2-x}C_x$. Solid line indicate the expected variation due to change in a-lattice parameter based on mode Gruineisen parameter reported in [25]. (b) Variation in the full width at half maximum, $\gamma$ with x in $MgB_{2-x}C_x$.

Fig. 4 Variation of $\gamma/\omega^2$ as a function of carbon fraction x in $MgB_{2-x}C_x$. Calculated variation [16] in the $\sigma$-hole density with x is shown as a solid line.

Fig. 5 Variation of $\lambda_{2g}$ with carbon fraction in $MgB_{2-x}C_x$.

Fig. 6 $T_C$ dependence on x in $MgB_{2-x}C_x$ calculated from the $\lambda_{2g}$ variation obtained from Raman data, and from the Hopfield expression. Experimentally measured $T_C$ is also shown for comparison.

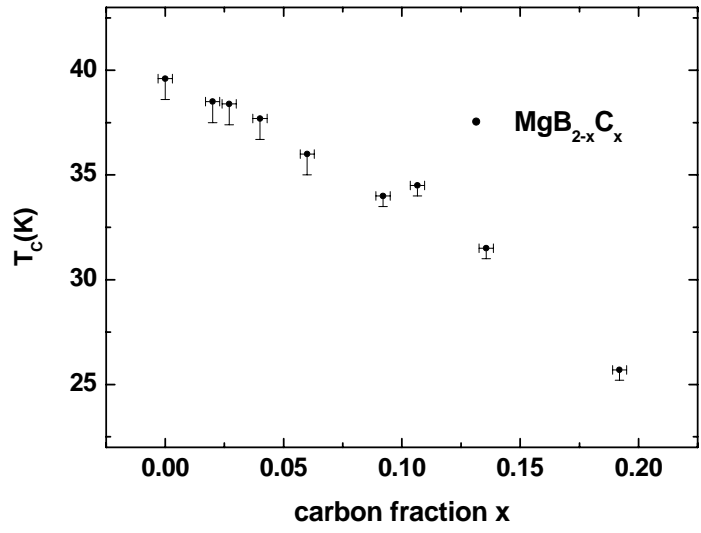

Fig. 1                    Sakuntala *et al*

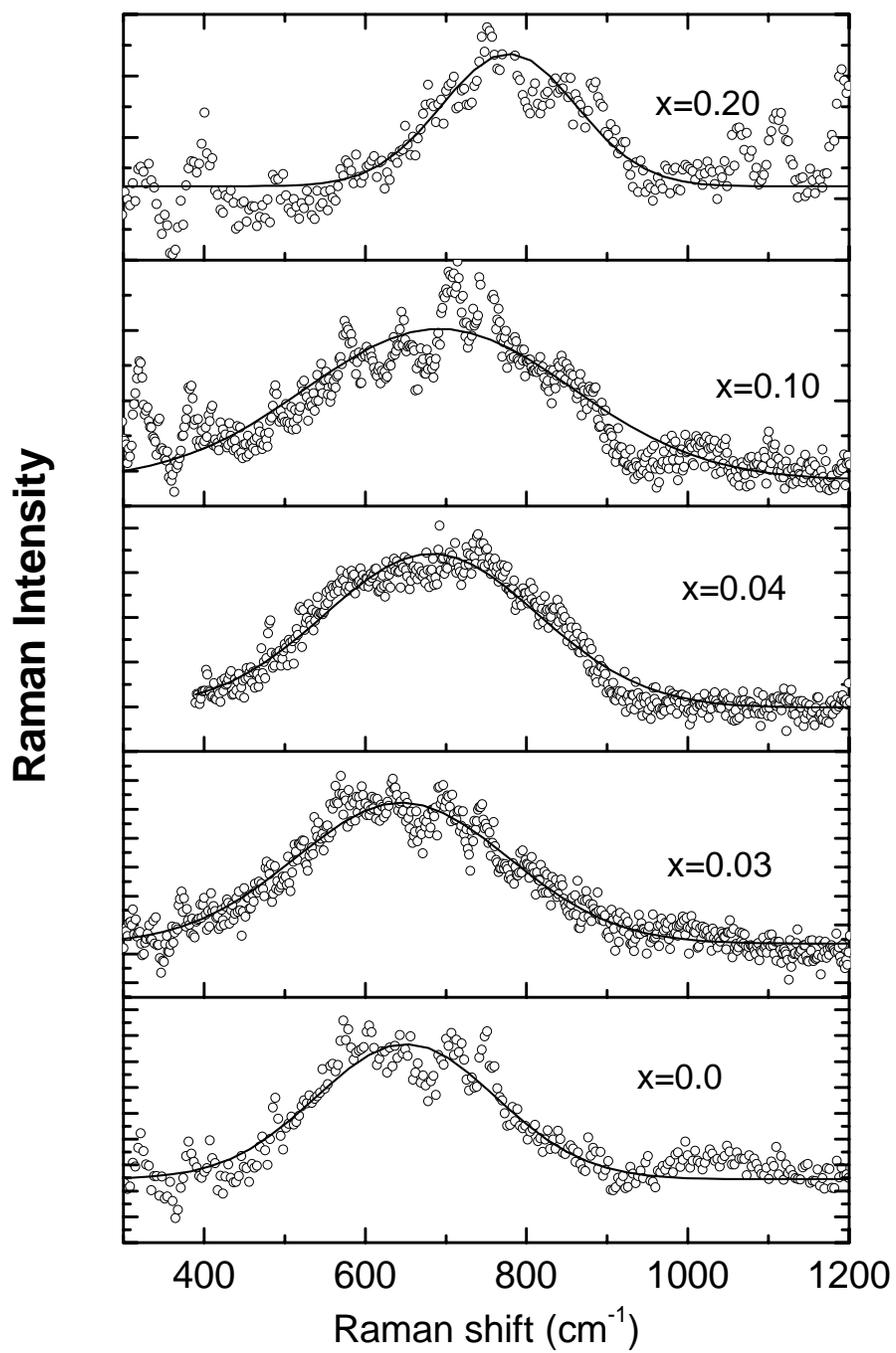

Fig. 2                    Sakuntala *et al*

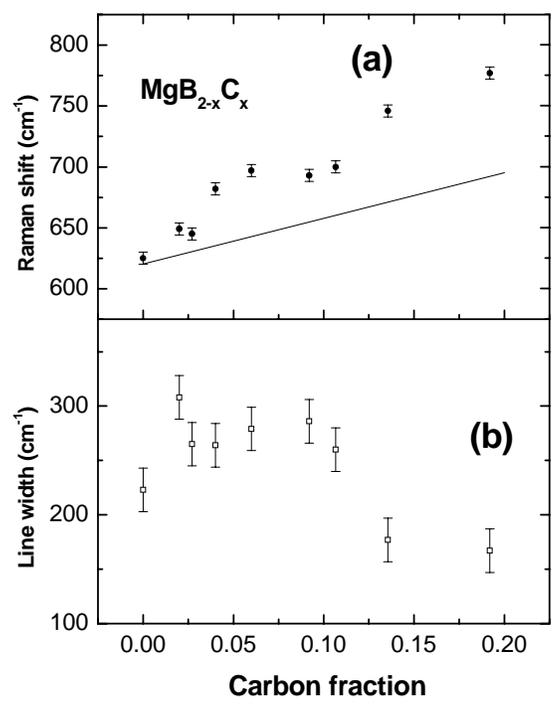

Fig. 3    Sakuntala *et al*

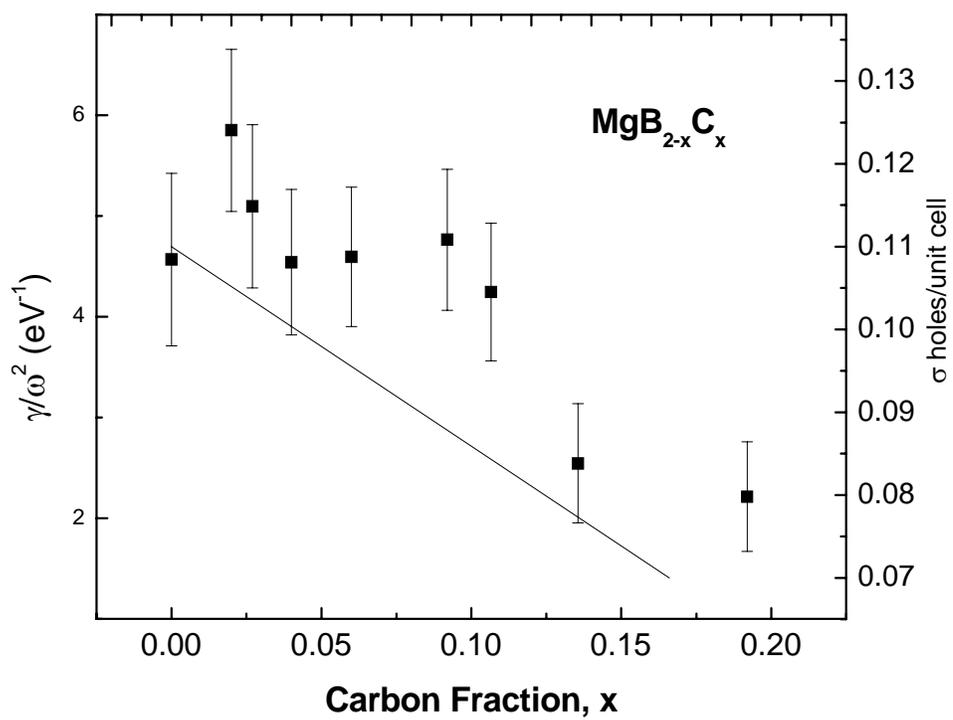

Fig. 4    Sakuntala *et al*

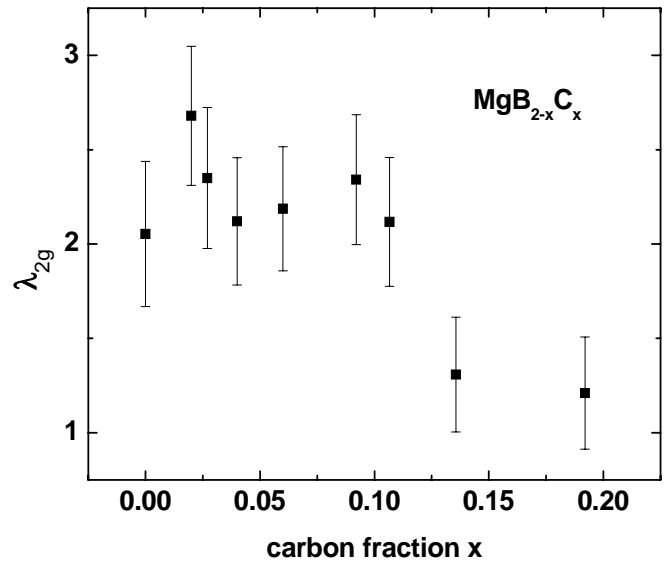

Fig. 5                           Sakuntala *et al*

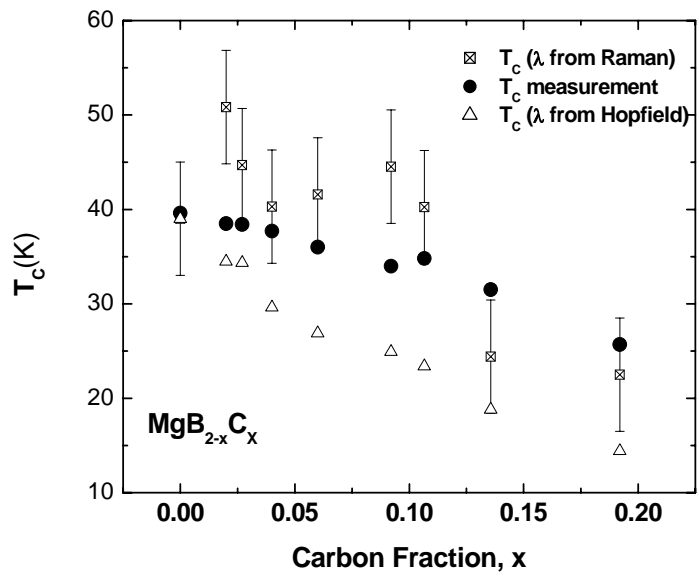

Fig. 6                    Sakuntala *et al*